\documentclass[conference]{IEEEtran}
\usepackage{amsmath,amssymb,amsfonts}
\usepackage{amsthm}
\usepackage{algorithmic}
\usepackage{algorithm}
\usepackage{array}
\usepackage{booktabs}
\usepackage{color}
\usepackage{cite}
\usepackage{graphicx}
\usepackage[colorlinks,linkcolor=red]{hyperref}
\usepackage{mathrsfs}
\usepackage{multirow}
\usepackage[caption=false,font=scriptsize,labelfont=sf,textfont=sf]{subfig}
\usepackage{textcomp}
\usepackage{threeparttable}
\usepackage{stfloats}
\usepackage{url}
\usepackage{verbatim}

\begin{document}

\title{Message-Enhanced DeGroot Model}

\author{\IEEEauthorblockN{Huisheng Wang, Zhanjiang Chen and H. Vicky Zhao}
\IEEEauthorblockA{\textit{Department of Automation, Tsinghua University}, Beijing, China\\
\url{whs22@mails.tsinghua.edu.cn}, \url{czj17@mails.tsinghua.edu.cn}, \url{vzhao@tsinghua.edu.cn}}}

\maketitle

\begin{abstract}
Understanding the impact of messages on agents' opinions over social networks is important. However, to our best knowledge, there has been limited quantitative investigation into this phenomenon in the prior works. To address this gap, this paper proposes the Message-Enhanced DeGroot model. The Bounded Brownian Message model provides a quantitative description of the message evolution, jointly considering temporal continuity, randomness, and polarization from mass media theory. The Message-Enhanced DeGroot model, combining the Bounded Brownian Message model with the traditional DeGroot model, quantitatively describes the evolution of agents' opinions under the influence of messages. We theoretically study the probability distribution and statistics of the messages and agents' opinions and quantitatively analyze the impact of messages on opinions. We also conduct simulations to validate our analyses.
\end{abstract}


\newcommand{\dif}{\mathrm{d}}
\newtheorem{definition}{\bf Definition}
\newtheorem{lemma}{\bf Lemma}
\newtheorem{theorem}{\bf Theorem}
\newtheorem{remark}{\bf Remark}

\section{Introduction}
Social media platforms, such as newspapers and broadcasting, serve as channels for disseminating messages and wielding substantial influence over agents' opinions \cite{guille2013information}. However, the spread of fake messages can potentially steer agents' opinions in unfavorable directions. It is essential to understand how messages shape agents' opinions, and effective guidance of agents' opinions has become a critical area of research \cite{burstein2003impact}.

From mass media theory, \textit{message} refers to specific information or content, while \textit{opinion} pertains to the subjective views and attitudes of agents towards the messages \cite{mcquail2010mcquail}. The prior work in \cite{lazarsfeld1968people} qualitatively showed that agents' opinions are influenced by both messages and the opinions of their neighbors, and both messages and agents' opinions evolve over time. Opinion dynamics offer valuable quantitative analysis tools for modeling the evolution of agents' opinions under the influence of the opinions of their neighbors over social networks \cite{noorazar2020recent}.  Classic models include the DeGroot model \cite{degroot1974reaching}, the Friedkin-Johnsen model \cite{friedkin1990social}, and others. The prior works in \cite{carletti2006make, hegselmann2006truth, gargiulo2008saturation, martins2010mass, kurz2011hegselmann, sirbu2013cohesion, sirbu2013opinion, li2020effect, yang2021opinion, muslim2024mass} analyzed the impact of messages on agents' opinions using opinion dynamics, while they assumed that messages remain static and do not change over time. The prior works in \cite{mandyam2012community, mirtabatabaei2014eulerian, quattrociocchi2014opinion, mao2018spread, mao2022social, gunducc2020effect} considered the scenarios where messages and agents' opinions are both dynamic, yet they assumed a simplified form for the message evolution, such as pure randomness or predetermined probability distributions. These assumptions of message evolution are simple, whereas message evolution in the real world is complex.

% \begin{figure}[!t]
% \centerline{\includegraphics[width=0.7\columnwidth]{1.pdf}}
% \caption{Message evolution and agents' opinions over social networks.}
% \label{fig:fig1}
% \end{figure}

From mass media theory, the message evolution exhibits three features: temporal continuity \cite{galtung1965structure}, randomness \cite{glasauer2022individual}, and polarization \cite{andina2007reinforcement, bernhardt2008political}. When a valuable topic is disseminated over social networks, it continues to attract agents' attention. Consequently, the messages related to this topic form a sequentially evolving sequence over time exhibiting \textit{temporal continuity}, and therefore, messages at different times are correlated \cite{galtung1965structure}. The \textit{randomness} refers to the uncertainty in the topic progression and the noise introduced by message sources when they edit the message contents \cite{glasauer2022individual}. Finally, when a message source adopts an extreme stance, it persistently emphasizes or reinforces that position in subsequent coverage, rather than shifting towards the opposite direction, which is referred to as \textit{polarization} \cite{andina2007reinforcement, bernhardt2008political}. To our knowledge, few works have quantitatively modeled the message evolution considering these three features and subsequently analyzed the impact of message evolution on opinion dynamics.

Taking into account the above features from mass media theory, in this paper, we quantitatively model the message evolution and incorporate it into the opinion dynamics model to study the impact of messages on agents' opinions. As an example, we study the impact of messages about stock prices on agents' opinions regarding the trend in the financial market.
From the prior work in \cite{osborne1959brownian}, the temporal continuity and randomness features of stock prices align with those of messages, and the feature that stock prices hit upper or lower limits is akin to polarization. 
In finance theory, Brownian motion has been extensively used to model the dynamic changes in stock prices and has been validated using real-world data \cite{reddy2016simulating}. Therefore, we use the Brownian motion to model the message evolution about stock prices. To model the polarization, we introduce two absorbing bounds into the Brownian motion model, which we call the \textit{Bounded Brownian Message} (BBM) model. 
Following the prior works in \cite{mao2018spread, mao2022social}, we model the messages and agents' opinions over social networks as a dynamic linear system, where agents' opinions are the state variables, while messages are the input variables. We incorporate the BBM model into the classic DeGroot opinion dynamics model \cite{degroot1974reaching} and propose the \textit{Message-Enhanced DeGroot} (MED) model to analyze the message and opinion evolution and study the impact of messages on agents' opinions.

Our contributions can be summarized as follows.
\begin{itemize}
    \item We propose the BBM model to model the message evolution, which jointly considers the temporal continuity, randomness, and polarization features, and we quantitatively analyze the statistics of messages.
    \item We propose the MED model to model the evolution of agents' opinions under the impact of messages, and quantitatively analyze the statistics of agents' opinions and the impact of messages on opinion dynamics.
    \item We conduct simulations to validate our analyses.
\end{itemize}

\section{The Bounded Brownian Message Model}
\label{sec:bbm}

In this section, we first introduce the BBM model and then analyze the statistical properties of the message, including its probability distribution and statistics (mean and variance). All detailed deviations and proofs are in [Appendix].

\subsection{Model Definition}
Assume that there are a total of $M$ message sources. We use $\mathscr{M}:=\{1,\dots,M\}$ to represent the index set of the message sources. Following the prior works in \cite{mirtabatabaei2014eulerian, mao2018spread, mao2022social}, we model the messages at time $t$ using a random vector $\boldsymbol{s}_t\in[0,1]$. Let $0$ and $1$ represent two messages supporting opposite and extreme views on a topic, which we call the two \textit{absorbing bounds}. In the BBM model, the message evolves following a stochastic process, and once it reaches one of the absorbing bounds, it remains unchanged. Let $s_{i,0}$ denote the initial message of the $i$-th message source. We assume that it follows a uniform distribution on $[\underline{\xi}, \overline{\xi}]\subset(0,1)$, and all initial messages are independent. The two bounds $\underline{\xi}$ and $\overline{\xi}$ prevent $s_{i,0}$ from approaching values of $0$ or $1$ quickly, avoiding an immediate convergence to the absorbing bounds.

To model the continuous-time message evolution, we assume that each message $\{s_{i,t}\}_{t\geqslant0}$ follows a Brownian motion. Let $(\Omega,\mathscr{F},\{\mathscr{F}_t\}_{t\geqslant0},\mathbb{P})$ be a complete filtered probability space on which an $M$-dimensional standard Brownian motion $\{\boldsymbol{z}_t\}_{t\geqslant0}$ is defined. The $i$-th component $\{z_{i,t}\}_{t\geqslant0}$ represents the standard Brownian motion for the $i$-th message source, and we assume that all components are independent. Let $c>0$ denote the changing rate of messages, which quantifies the level of randomness, and we assume that $c$ is identical for all messages. Without considering the absorbing bounds, the message of the $i$-th message source at time $t$ is
\begin{equation}
    y_{i,t}=s_{i,0}+cz_{i,t}.
    \label{eq:bbm0}
\end{equation}
When we consider the absorbing bounds, i.e., the message of the $i$-th message source stops changing once it reaches $0$ or $1$, the message at time $t$ can be expressed as
\begin{equation}
    s_{i,t}=y_{i,u},
    \label{eq:bbm}
\end{equation}
where $u:=t\wedge T_{i,0}\wedge T_{i,1}$. Here, we define $T_{i,0}$ as the first time $y_{i,t}$ hits $0$, i.e., $T_{i,0} := \inf\{t : t > 0, y_{i,t} = 0\}$, and $T_{i,1}$ is defined similarly. $\wedge$ returns the infimum of two variables. Equation \eqref{eq:bbm0} and \eqref{eq:bbm} are called the BBM model. 

In the BBM model, the message is a continuous-time stochastic process, which reflects the time continuity feature of the message. The Brownian motion term reflects the randomness feature. Additionally, the two absorbing bounds in the BBM model reflect the polarization feature. Therefore, the BBM model can quantitatively describe the complex characteristics of messages in mass media theory.

\subsection{Probability
Distribution and Statistics of the Message}
Given the BBM model, we can analyze the probability distribution and statistics of the message.

\subsubsection{Probability distribution of the message}
Theorem \ref{th:1} gives the probability distribution of $s_{i,t}$.

\begin{theorem}\label{th:1}
The conditional probabilities of the message for the $i$-th message source that starts with $s_{i,0}$ and reaches the absorbing bounds $0$ and $1$ at time $t$ are
\begin{equation}
\begin{cases}
    \displaystyle \mathbb{P}(s_{i,t}=0|s_{i,0}) = \int_0^t\frac{\dif \tau}{\tau}\sum_{j\in\mathbf{E}}(s_{i,0}-j)g(j,\tau|s_{i,0}),\\
    \displaystyle \mathbb{P}(s_{i,t}=1|s_{i,0}) = \int_0^t\frac{\dif \tau}{\tau}\sum_{j\in\mathbf{O}}(j-s_{i,0})g(j,\tau|s_{i,0}),\label{equ:si=0}
\end{cases}
\end{equation}
where $\mathbf{E}$ and $\mathbf{O}$ refer to the sets of the even and odd numbers, and $g(x,t|x_0):=\frac{1}{\sqrt{2\pi c^2t}}\exp\left\{-\frac{(x-x_0)^2}{2c^2t}\right\}$ is the transition probability from $x_0$ to $x$ over time $t$  of the Brownian motion with changing rate $c$. The conditional probability density function of $s_{i,t}\in(0,1)$ is
\begin{equation}
f_{s_{i,t}}(x|s_{i,0}) = \sum_{n\in\mathbf{E}}g(n+x,t|s_{i,0})-g(-n-x,t|s_{i,0}).
\label{eq:pdf}
\end{equation}
\end{theorem}

From Theorem \ref{th:1}, we can show that when $s_{i,0} <\frac{1}{2}$, $\mathbb{P}(s_{i,t} = 0|s_{i,0}) > \mathbb{P}(s_{i,t} = 1|s_{i,0})$, and conversely, when $s_{i,0} > \frac{1}{2}$, $\mathbb{P}(s_{i,t} = 0|s_{i,0}) < \mathbb{P}(s_{i,t} = 1|s_{i,0})$. This indicates that if the initial message leans towards one side, then as the message evolves, the distribution will also tend towards that side. 
Furthermore, the probability of the message reaching the absorbing bounds increases monotonically with time, implying that as the message evolves, it tends to become more polarized.  It can be proved that as $t$ approaches infinity, the message almost surely reaches the absorbing bounds, and adheres to a binomial distribution, as stated in Theorem \ref{co:1}.
\begin{theorem}\label{co:1}
    As $t$ approaches infinity, the probability distribution of $s_{i,t}$ is a binomial distribution, and
    \begin{equation}
    \begin{cases}
        \displaystyle\lim_{t\to\infty}\mathbb{P}(s_{i,t}=1|s_{i,0})\overset{\text{a.s.}}{=}s_{i,0},\\
        \displaystyle\lim_{t\to\infty}\mathbb{P}(s_{i,t}=0|s_{i,0})\overset{\text{a.s.}}{=}1 - s_{i,0},\\
        \displaystyle\lim_{t\to\infty}f_{s_{i,t}}(x|s_{i,0})\overset{\text{a.s.}}{=}0,\quad\forall x\in(0, 1).
    \end{cases}
    \label{eq:asy}
    \end{equation}
\end{theorem}
Here, $\overset{\text{a.s.}}{=}$ represents the two variables that are equal in probability almost surely. And we define $\overset{\text{a.s.}}{\leqslant}$ similarly.

\subsubsection{Statistics of the message}
In the BBM model, the message is a stochastic process. With the probability distribution of the message, we can further analyze its statistics. First, we analyze the mean of the message. 

\begin{theorem}\label{th:2}
The conditional mean of $s_{i,t}$ given $s_{i,0}$ is
\begin{equation}
    \mathbb{E}(s_{i,t}|s_{i,0})\overset{\text{a.s.}}{=}s_{i,0},
    \label{eq:ess}
\end{equation}
and the mean of $s_{i,t}$ is
    \begin{equation}
        \mathbb{E}s_{i,t}=\mu,
        \label{eq:mean}
    \end{equation}
    where $\mu:=\frac{1}{2}(\underline{\xi}+\overline{\xi})$ is the mean of the uniform distribution on $[\underline{\xi}, \overline{\xi}]$. In vector form, \eqref{eq:mean} is
    \begin{equation}
        \mathbb{E}\boldsymbol{s}_t=\mu\boldsymbol{1},
        \label{eq:mean-vec}
    \end{equation}
    where $\boldsymbol{1}$ is the all one vector.
\end{theorem}

From Theorem \ref{th:2}, the mean of the message remains the same over time and equals the mean of the initial message distribution. 

Next, we analyze the variance of the message. It is complicated to calculate the exact value of the variance of $s_{i,t}$, and we show its upper bound in Theorem \ref{th:3}.
\begin{theorem}\label{th:3}
    The upper bound of the conditional variance of $s_{i,t}$ given $s_{i,0}$ is
    \begin{equation}
        \mathbb{D}(s_{i,t}|s_{i,0})\overset{\text{a.s.}}{\leqslant}(c^2t)\wedge[s_{i,0}(1-s_{i,0})],
        \label{eq:dss}
    \end{equation}
    and the upper bound of variance of $s_{i,t}$ is
    \begin{equation}
        \mathbb{D}s_{i,t}\leqslant(c^2t+\delta^2)\wedge[\mu(1-\mu)],
        \label{eq:var}
    \end{equation}
    where $\delta^2:=\frac{1}{12}(\overline{\xi}-\underline{\xi})^2$ is the variance of the uniform distribution in $[\underline{\xi}, \overline{\xi}]$. In vector form, \eqref{eq:var} is
    \begin{equation}
        \mathbb{D}\boldsymbol{s}_t\leqslant(c^2t+\delta^2)\wedge[\mu(1-\mu)]\boldsymbol{1}.
        \label{eq:var-vec}
    \end{equation}
\end{theorem}

From Theorem \ref{th:3}, the upper bound of the variance is determined by two terms, each corresponding to the following two scenarios. When $t$ is small, the message has not yet reached the absorbing bounds, and the stochastic process still exhibits Brownian motion features, with its conditional variance increasing linearly with time, i.e.,
\begin{equation}
    \mathbb{D}s_{i,t}=c^2t+\delta^2 \ \text{for small value of} \ t.
\end{equation}
As $t$ approaches infinity, the message almost surely reaches the absorbing bound, and the distribution of the message takes on a binomial distribution, as described in Theorem \ref{co:1}. In this case, the variance reaches a steady state, i.e.,
\begin{equation}
    \lim_{t\to\infty}\mathbb{D}s_{i,t}=\mu(1-\mu).
    \label{eq:vars}
\end{equation}

Furthermore, it can be proved that the variance of the message is a monotonically increasing function with time.

In summary, based on Theorem \ref{th:1}--\ref{th:3}, in the BBM model, all messages initially follow a uniform distribution on a proper subinterval of $[0,1]$, and eventually converge to a $0/1$ binomial distribution. Throughout the message evolution, the mean remains constant while the variance gradually increases. These properties align with the features of time continuity, randomness, and polarization in mass media theory.

\section{The Message-Enhanced DeGroot Model}
\label{sec:med}
The BBM model models the message evolution, which jointly considers the temporal continuity, randomness, and polarization features. Next, we incorporate the BBM model into the traditional DeGroot opinion dynamics model and propose the MED model. Subsequently, we quantitatively analyze the agents' opinions dynamics over social networks and the impact of messages on agents' opinions based on Theorem \ref{th:1}--\ref{th:3}. All detailed deviations and proofs are in [Appendix].

\subsection{The Traditional DeGroot Model}
Assume that there are a total of $N$ agents in a fully connected network. We use $\mathscr{N}:=\{1,\dots,N\}$ to represent the index set of the agents. Let $\boldsymbol{W}\in\mathbf{R}^{N\times N}$ denote the adjacency matrix, which is a stochastic matrix with $\sum_{j=1}^Nw_{ij}=1,\forall i\in\mathscr{N}$ and $w_{ij}\geqslant0,\forall(i,j )\in\mathscr{N}\times\mathscr{N}$. $w_{ij}$ quantifies the $j$-th agent's influence on the $i$-th agent in opinion update. 

Let $\boldsymbol{o}_t$ denote the vector of agents' opinion at time $t$. Given the initial opinion $\boldsymbol{o}_0$, the continuous-time form of the DeGroot model is described by the following differential equation \cite{okawa2022predicting}:
\begin{equation}
    \dot{\boldsymbol{o}}_t=(\boldsymbol{W}-\boldsymbol{I})\boldsymbol{o}_t,
    \label{eq:dg}
\end{equation}
where $\boldsymbol{I}\in\mathbf{R}^{N\times N}$ is the identity matrix. In \eqref{eq:dg}, the changing rate of agents' opinions $\dot{\boldsymbol{o}}_t$ at time $t$ is a linear combination of the agents' opinions $\boldsymbol{o}_t$ at time $t$, and the opinions are only determined by the adjacency matrix $\boldsymbol{W}$ and initial opinion $\boldsymbol{o}_0$ without considering the external messages.

From \cite{berger1981necessary}, the opinion at time $t$ is
\begin{equation}
    \boldsymbol{o}_t=\mathrm{e}^{(\boldsymbol{W}-\boldsymbol{I})t}\boldsymbol{o}_0,
\end{equation}
where $\boldsymbol{o}_0$ is the vector of agents' initial opinions, and $\mathrm{e}^{(\boldsymbol{W}-\boldsymbol{I})t}:=\sum_{k=0}^{\infty}\frac{t^k}{k!}(\boldsymbol{W}-\boldsymbol{I})^k$. As $t$ approaches infinity, the \textit{steady-state opinion} is
\begin{equation}
    \lim_{t\to\infty}\boldsymbol{o}_t=\boldsymbol{\ell}^\top\boldsymbol{o}_0\boldsymbol{1},
    \label{eq:dg-o}
\end{equation}
where $\boldsymbol{\ell}\in\mathbf{R}^N$ is the eigenvector of $\boldsymbol{W}$ associated with $1$ constrained to $\sum_{i=1}^N\ell_i=1$. 

\subsection{The Message-Enhanced DeGroot Model}
\subsubsection{Model definition}
Following the prior works in \cite{mao2018spread, yang2021opinion, mao2022social}, we model the messages
and agents' opinions as a dynamic system, where
agents' opinions are the state variables, while messages are the input variables. The incorporation of messages described by the BBM model with the DeGroot model leads us to the following differential equation:
\begin{equation}
    \dot{\boldsymbol{o}}_t=(\alpha\boldsymbol{W}-\boldsymbol{I})\boldsymbol{o}_t+(1-\alpha)\boldsymbol{U}\boldsymbol{s}_t,
    \label{eq:med}
\end{equation}
where $\alpha\in(0, 1)$ is the weight coefficient of the opinion, and $1-\alpha$ measures the extent to which the messages affect the opinion update. A smaller $\alpha$ indicates that messages exert a greater impact on the changing rate of agents' opinions. When $\alpha = 1$, the MED model degenerates into the traditional DeGroot model in \eqref{eq:dg}. The influence matrix $\boldsymbol{U}\in\mathbf{R}^{N\times M}$ is a stochastic matrix with $\sum_{j=1}^Mu_{ij}=1,\forall i\in\mathscr{N}$ and $u_{ij}\geqslant0,\forall(i,j)\in\mathscr{N}\times\mathscr{M}$. $u_{ij}$ quantifies the influence of the $j$-th source on the $i$-th agent in his/her opinion update. 
Equation \eqref{eq:med} is called the MED model.

In the MED model, the changing rate in opinions $\dot{\boldsymbol{o}}_t$ at time $t$ is composed of two parts: the linear combination of opinions from all agents $(\alpha\boldsymbol{W}-\boldsymbol{I})\boldsymbol{o}_t$ at time $t$ and the linear combination of messages $(1-\alpha)\boldsymbol{U}\boldsymbol{s}_t$ at time $t$. 
We can calculate the opinion at time $t$ from \eqref{eq:med}. 
\begin{theorem}\label{th:4}
    The opinion $\boldsymbol{o}_t$ at time $t$ is
    \begin{equation}
    \boldsymbol{o}_t=\mathrm{e}^{(\alpha\boldsymbol{W}-\boldsymbol{I})t}\boldsymbol{o}_0+(1-\alpha)\int_0^t\mathrm{e}^{(\alpha\boldsymbol{W}-\boldsymbol{I})(t-\tau)}\boldsymbol{U}\boldsymbol{s}_{\tau}\dif\tau.
\end{equation}
\end{theorem}

From Theorem \ref{th:4}, agents' opinions are influenced not only by the adjacency matrix $\boldsymbol{W}$ and the initial opinions $\boldsymbol{o}_0$ but also by the messages $\{\boldsymbol{s}_t\}_{t\geqslant0}$ and the influence matrix $\boldsymbol{U}$. 

\subsubsection{Statistics of the opinion}
From Theorem \ref{th:2}--\ref{th:4}, we analyze the mean and variance of agents' opinions in the MED model.

\begin{theorem}\label{co:4}
    The mean of $\boldsymbol{o}_t$ is
\begin{equation}
    \mathbb{E}\boldsymbol{o}_t=\mathrm{e}^{(\alpha\boldsymbol{W}-\boldsymbol{I})t}\boldsymbol{o}_0+\mu\left[\boldsymbol{I}-\mathrm{e}^{(\alpha\boldsymbol{W}-\boldsymbol{I})t}\right]\boldsymbol{1}.
    \label{eq:Eo(t)}
\end{equation}
As $t$ approaches infinity,
\begin{equation}
    \lim_{t\to\infty}\mathbb{E}\boldsymbol{o}_t=\mu\boldsymbol{1}.
    \label{eq:lime}
\end{equation}
\end{theorem}

From Theorem \ref{co:4}, $\mathbb{E}\boldsymbol{o}_t$ transitions from $\boldsymbol{o}_0$ to $\mu\boldsymbol{1}$ over time, and as $t$ approaches infinity, $\mathbb{E}\boldsymbol{o}_t$
converges to the mean of the message distribution.

% \begin{theorem}\label{co:5}
% The variance of $\boldsymbol{o}_t$ is
% \begin{align}
%     \mathbb{D}\boldsymbol{o}_t&=(1-\alpha)^2\mathrm{diag}\left[\int_0^t\int_0^t\mathrm{e}^{(\alpha\boldsymbol{W}-\boldsymbol{I})(t-\sigma)}\right.\notag\\
%     &\cdot\left.\boldsymbol{U}\boldsymbol{\Sigma}_{\sigma,\tau}\boldsymbol{U}^\top \mathrm{e}^{(\alpha\boldsymbol{W}^\top-\boldsymbol{I})(t-\tau)}\dif \sigma\dif\tau\right],
%     \label{eq:Do(t)1}
% \end{align}
% where $\boldsymbol{\Sigma}_{\sigma,\tau}:=\mathrm{Cov}(\boldsymbol{s}_\sigma,\boldsymbol{s}_\tau)$ is the autocovariance matrix of the messages, and $\mathrm{diag}$ transforms the diagonal of a matrix into a column vector. As $t$ approaches infinity,
%     \begin{align}
%         \lim_{t\to\infty}\mathbb{D}\boldsymbol{o}_t&=\mu(1-\mu)(1-\alpha)^2\notag\\
%         &\cdot\mathrm{diag}[(\alpha\boldsymbol{W}-\boldsymbol{I})^{-1}\boldsymbol{U}\boldsymbol{U}^\top (\alpha\boldsymbol{W}^\top-\boldsymbol{I})^{-1}].
%         \label{eq:Do(t)2}
%     \end{align}
% \end{theorem}

Because it is difficult to analyze $\mathbb{D}\boldsymbol{o}_t$ for each $t$, we focus on the variance of agents' opinions as $t$ approaches infinity. 
\begin{theorem}\label{co:5}
As $t$ approaches infinity, the variance of $\boldsymbol{o}_t$ is
    \begin{align}
        \lim_{t\to\infty}\mathbb{D}\boldsymbol{o}_t&=\mu(1-\mu)(1-\alpha)^2\notag\\
        &\cdot\mathrm{diag}[(\alpha\boldsymbol{W}-\boldsymbol{I})^{-1}\boldsymbol{U}\boldsymbol{U}^\top (\alpha\boldsymbol{W}^\top-\boldsymbol{I})^{-1}],
        \label{eq:Do(t)2}
    \end{align}
    where $\mathrm{diag}$ transforms the diagonal of a matrix into a vector.
\end{theorem}

From \eqref{eq:Do(t)2}, as $t$ approaches infinity, the variance of agents' opinions is proportional to the variance of messages $\mu(1-\mu)$, as shown in \eqref{eq:vars}. The vector $(1-\alpha)^2\cdot\mathrm{diag}[(\alpha\boldsymbol{W}-\boldsymbol{I})^{-1}\boldsymbol{U}\boldsymbol{U}^\top (\alpha\boldsymbol{W}^\top-\boldsymbol{I})^{-1}]$ is the weighted average coefficient of all messages. 

% The matrix $\boldsymbol{B}$ quantifies the direct impact of messages on agents' opinions. Through the matrix $\boldsymbol{A}$, the impact of messages on opinions is manifested in the mutual interactions among agents' opinions. In this context, the matrix $\boldsymbol{A}$ characterizes the indirect impact of messages on agents' opinions.

% For larger values of $t$, the variance of $o_{i,t}$ is proportional to the $i$-th element of the diagonal of $\boldsymbol{A}^{-1}\boldsymbol{B}\boldsymbol{B}^\top \boldsymbol{A}^{-\top}$. This diagonal represents the sum of squared message influences, weighted by group interactions. In summary, when an agent's acceptance of all messages is roughly equal, their opinion's variance approaches decrease. This is because when message influences are similar, it is akin to averaging all messages, reducing the uncertainty of each individual message. Additionally, with an increase in the number of message sources, the uncertainty in the average diminishes.

\subsubsection{Impact of the Message on the Opinion}
Next, we compare the MED model with the DeGroot model and study the impact of the messages on agents' opinions. 

First, due to the randomness of messages, agents' opinions in the MED model are a stochastic process. However, in the DeGroot model, the vector of the opinions is deterministic given the initial opinions and the adjacency matrix. 

Second, as shown in \eqref{eq:Eo(t)}, the mean of agents' opinions consists of two terms in the MED model. The first term quantifies the impact of the initial opinions and the adjacency matrix, which is similar to the DeGroot model, while the second term quantifies the impact of messages. As the time approaches infinity, the mean of agents' steady-state opinions converges to the mean of the message distribution. However, as shown in \eqref{eq:dg-o}, in the DeGroot model, agents' steady-state opinions are determined by the initial opinions and the adjacency matrix.

Third, in the MED model, the variance of agents' opinions increases over time, which is attributed to the growing uncertainty in the messages. In contrast, the DeGroot model's opinions are deterministic, so the variance is always zero. 

In summary, in the BBM model, external messages cause stochastic fluctuations in agents' opinions over time, leading to increasing variance. Additionally, the mean of agents' steady-state opinions in the BBM model converges to the mean of the message distribution, reflecting the influence of external messages on opinion dynamics.

\section{Simulation Analysis}
\label{sec:experiments}

% \begin{figure*}[!t]
% \centering
% \subfloat[]{\includegraphics[width=0.2\linewidth]{fig2.pdf} \label{fig:fig2a}}
% \subfloat[]{\includegraphics[width=0.2\linewidth]{fig3.pdf} \label{fig:fig2b}}
% \subfloat[]{\includegraphics[width=0.2\linewidth]{fig4.pdf} \label{fig:fig2c}}
% \subfloat[]{\includegraphics[width=0.2\linewidth]{fig5.pdf} \label{fig:fig2d}}
% \hfill
% \subfloat[]{\includegraphics[width=0.2\linewidth]{fig6.pdf} \label{fig:fig2e}}
% \subfloat[]{\includegraphics[width=0.2\linewidth]{fig7.pdf} \label{fig:fig2f}}
% \subfloat[]{\includegraphics[width=0.2\linewidth]{fig8.pdf} \label{fig:fig2g}}
% \subfloat[]{\includegraphics[width=0.2\linewidth]{fig9.pdf} \label{fig:fig2h}}
% \caption{Simulation results.}
% \label{fig:fig2}
% \end{figure*}

In this section, we conduct simulations to verify the analyses in Section \ref{sec:bbm} and \ref{sec:med}.

\subsection{The BBM model}

\begin{figure}[!t]
\centering
\subfloat[Probability reaching $0$ and $1$]{\includegraphics[width=0.45\linewidth]{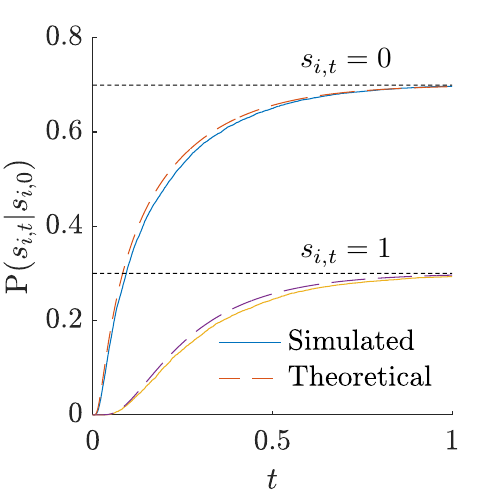} \label{fig:fig2a}}
\subfloat[Probability density function]{\includegraphics[width=0.45\linewidth]{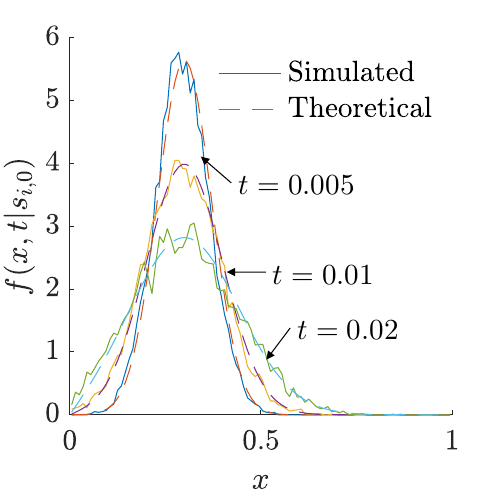} \label{fig:fig2b}}
\caption{Probability distribution of the message in the BBM model.}
\label{fig:fig2}
\end{figure}

\begin{figure}[!t]
\centering
\subfloat[Mean]{\includegraphics[width=0.45\linewidth]{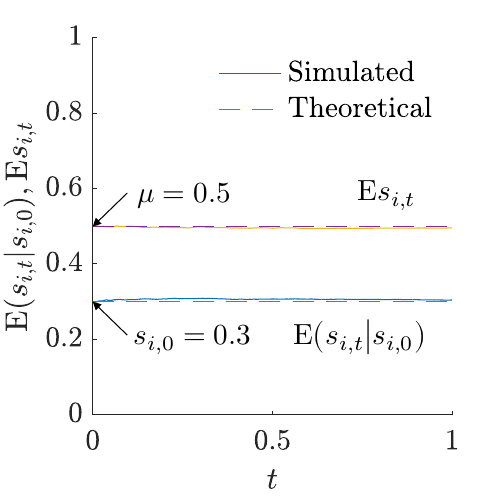} \label{fig:fig2c}}
\subfloat[Variance]{\includegraphics[width=0.45\linewidth]{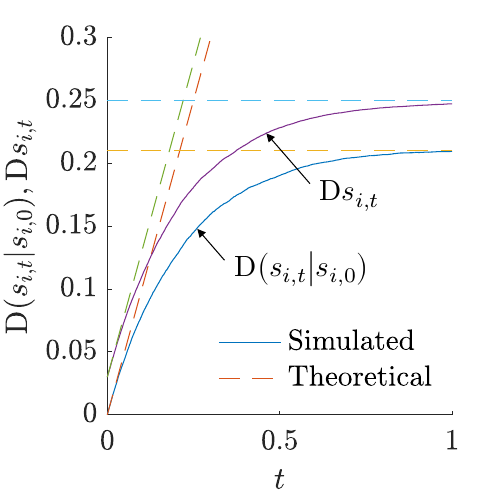} \label{fig:fig2d}}
\caption{Statistics of the message in the BBM model.}
\label{fig:fig3}
\end{figure}

We first simulate the message evolution and verify our analyses of the statistical properties of the message in Section \ref{sec:bbm}. We set the parameters as follows: $s_0 = 0.3$, $c = 1$. We generate a total of 10,000 trajectories of the stochastic process. 

\subsubsection{Probability distribution of the message}
The curves for $\mathbb{P}(s_t = 1 | s_0 = 0.3)$ and $\mathbb{P}(s_t = 0 | s_0 = 0.3)$ are shown in Fig. \ref{fig:fig2a}. Solid lines represent simulation values computed using the frequency of messages reaching the absorbing bounds at $0$ and $1$ at time $t$, respectively. Dashed lines represent theoretical values computed using \eqref{equ:si=0}. From Fig. \ref{fig:fig2a}, simulation results match our theoretical analysis results very well, validating the correctness of Theorem \ref{th:1}.
Fig. \ref{fig:fig2a} illustrates that the probability of a message reaching the absorbing bounds monotonically increases with time. The probability distribution functions $f_{s_{i,t}}(x| s_0 = 0.3)$ at $t = 0.005, 0.01, 0.02$ are plotted in Fig. \ref{fig:fig2b}. Again, the simulation and theoretical values exhibit a notable similarity, confirming the validity of Theorem \ref{th:1}.

From Fig. \ref{fig:fig2a}, as $t$ approaches infinity, $\lim_{t\to\infty}\mathbb{P}(s_t = 1 | s_0 = 0.3)=0.3$ and $\lim_{t\to\infty}\mathbb{P}(s_t = 0 | s_0 = 0.3)=0.7$, consistent with Theorem \ref{co:1}. 

\subsubsection{Statistics of the message}
We simulate the mean and variance of messages with $\underline{\xi} = 0.2$, $\overline{\xi} = 0.8$, $c=1$. When we calculate the conditional mean and conditional variance, we fix $s_{i,0}=0.3$. As shown in Fig. \ref{fig:fig3}, solid lines represent simulation values calculated by statistical methods, while dashed lines represent the theoretical values computed using \eqref{eq:ess}, \eqref{eq:mean}, \eqref{eq:dss}, and \eqref{eq:var}, respectively.
The simulation values closely align with theoretical values, validating Theorem \ref{th:2} and Theorem \ref{th:3}.

\subsection{The MED model}
\begin{figure}[!t]
\centering
\subfloat[Mean]{\includegraphics[width=0.45\linewidth]{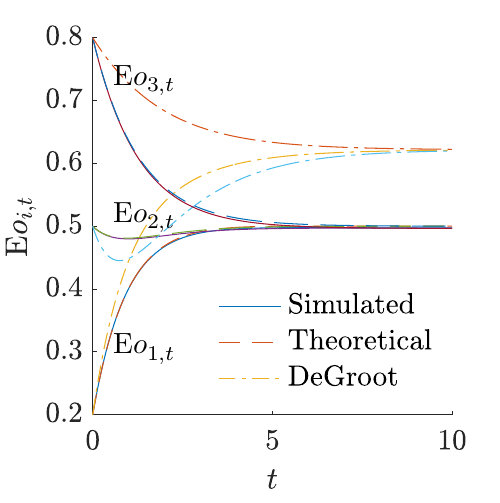} \label{fig:fig2g}}
\subfloat[Variance]{\includegraphics[width=0.45\linewidth]{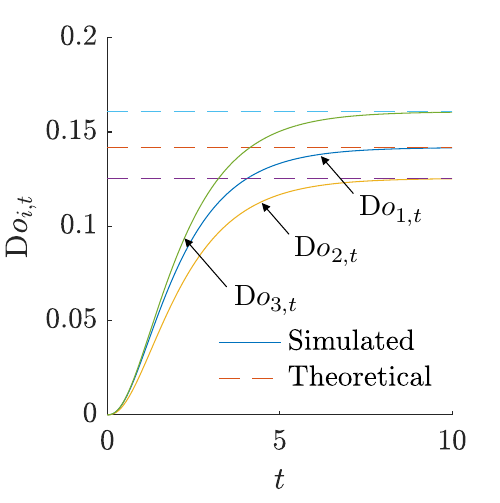} \label{fig:fig2h}}
\caption{Statistics of the opinion in the MED model.}
\label{fig:fig4}
\end{figure}

Next, we simulate the opinion dynamics under the impact of the messages and verify our analyses of the statistical properties of the opinion in Section \ref{sec:med}. We set the parameters as follows: $M=2$, $N=3$, $\underline{\xi} = 0.2$, $\overline{\xi} = 0.8$, $c=1$, $\boldsymbol{W}=\left[ \begin{smallmatrix}
0.2 & 0.3 & 0.2 \\
0.7 & 0.2 & 0.1 \\
0.1 & 0.1 & 0.8
\end{smallmatrix} \right]$, $\boldsymbol{U}=\left[ \begin{smallmatrix}
0.8 & 0.2 \\
0.2 & 0.2 \\
0.2 & 0.8
\end{smallmatrix} \right]$, $\alpha=0.3$, and $\boldsymbol{o}_0=\left[ \begin{smallmatrix}
0.2 \\
0.2 \\
0.8
\end{smallmatrix} \right]$. The experiment repeats 10,000 times, and the curves for the mean and variance of agents' opinions are plotted in Fig. \ref{fig:fig4}.

\subsubsection{Mean of the opinion}
In Fig. \ref{fig:fig2g}, solid lines represent the simulated mean of agents' opinions calculated using statistical methods, and dashed lines represent the theoretical mean of agents' opinions calculated using \eqref{eq:Eo(t)}. Simulation results match our theoretical analysis results very well. As the time approaches infinity, all agents' opinions converge to the mean of the initial message distribution, validating Theorem \ref{co:4}.

Given $\boldsymbol{W}$ and $\boldsymbol{o}_0$, the opinion curves based on the DeGroot model, with identical simulation parameters, are also included as dashed-dotted lines in Fig. \ref{fig:fig2g}. The steady-state opinions deviate from the mean of initial message distribution in the DeGroot model.

\subsubsection{Variance of the opinion}
In Fig. \ref{fig:fig2h}, solid lines represent the simulated variance of agents' opinions calculated using statistical methods. Dashed lines represent theoretical variance values calculated using \eqref{eq:Do(t)2}. Once again, a close correspondence between simulation and theoretical results validates the correctness of Theorem \ref{co:5}.

\section{Conclusions}
\label{sec:conclusion}
In this paper, we propose the BBM model to model the message evolution, and the MED model to study the opinion dynamics under the impact of messages over social networks. We theoretically study the probability distribution and statistics of the messages and agents' opinions and quantitatively analyze the impact of messages on agents' opinions. Simulation results validate our analyses.

\bibliographystyle{IEEEtran}
\bibliography{bibliology}

% Generated by IEEEtran.bst, version: 1.14 (2015/08/26)
\begin{thebibliography}{10}
\providecommand{\url}[1]{#1}
\csname url@samestyle\endcsname
\providecommand{\newblock}{\relax}
\providecommand{\bibinfo}[2]{#2}
\providecommand{\BIBentrySTDinterwordspacing}{\spaceskip=0pt\relax}
\providecommand{\BIBentryALTinterwordstretchfactor}{4}
\providecommand{\BIBentryALTinterwordspacing}{\spaceskip=\fontdimen2\font plus
\BIBentryALTinterwordstretchfactor\fontdimen3\font minus \fontdimen4\font\relax}
\providecommand{\BIBforeignlanguage}[2]{{%
\expandafter\ifx\csname l@#1\endcsname\relax
\typeout{** WARNING: IEEEtran.bst: No hyphenation pattern has been}%
\typeout{** loaded for the language `#1'. Using the pattern for}%
\typeout{** the default language instead.}%
\else
\language=\csname l@#1\endcsname
\fi
#2}}
\providecommand{\BIBdecl}{\relax}
\BIBdecl

\bibitem{guille2013information}
A.~Guille, H.~Hacid, C.~Favre, and D.~A. Zighed, ``Information diffusion in online social networks: A survey,'' \emph{ACM Sigmod Record}, vol.~42, no.~2, pp. 17--28, 2013.

\bibitem{burstein2003impact}
P.~Burstein, ``The impact of public opinion on public policy: {A} review and an agenda,'' \emph{Political research quarterly}, vol.~56, no.~1, pp. 29--40, 2003.

\bibitem{mcquail2010mcquail}
D.~McQuail, \emph{McQuail's mass communication theory}.\hskip 1em plus 0.5em minus 0.4em\relax Sage publications, 2010.

\bibitem{lazarsfeld1968people}
P.~F. Lazarsfeld, B.~Berelson, and H.~Gaudet, ``The people’s choice,'' in \emph{The people’s choice}.\hskip 1em plus 0.5em minus 0.4em\relax Columbia University Press, 1968.

\bibitem{noorazar2020recent}
H.~Noorazar, ``Recent advances in opinion propagation dynamics: {A} 2020 survey,'' \emph{The European Physical Journal Plus}, vol. 135, no.~6, pp. 1--20, 2020.

\bibitem{degroot1974reaching}
M.~H. DeGroot, ``Reaching a consensus,'' \emph{Journal of the American Statistical association}, vol.~69, no. 345, pp. 118--121, 1974.

\bibitem{friedkin1990social}
N.~E. Friedkin and E.~C. Johnsen, ``Social influence and opinions,'' \emph{Journal of Mathematical Sociology}, vol.~15, no. 3-4, pp. 193--206, 1990.

\bibitem{carletti2006make}
T.~Carletti, D.~Fanelli, S.~Grolli, and A.~Guarino, ``How to make an efficient propaganda,'' \emph{Europhysics Letters}, vol.~74, no.~2, p. 222, 2006.

\bibitem{hegselmann2006truth}
R.~Hegselmann, U.~Krause \emph{et~al.}, ``Truth and cognitive division of labor: First steps towards a computer aided social epistemology,'' \emph{Journal of Artificial Societies and Social Simulation}, vol.~9, no.~3, p.~10, 2006.

\bibitem{gargiulo2008saturation}
F.~Gargiulo, S.~Lottini, and A.~Mazzoni, ``The saturation threshold of public opinion: Are aggressive media campaigns always effective?'' \emph{arXiv preprint arXiv:0807.3937}, 2008.

\bibitem{martins2010mass}
T.~V. Martins, M.~Pineda, and R.~Toral, ``Mass media and repulsive interactions in continuous-opinion dynamics,'' \emph{Europhysics Letters}, vol.~91, no.~4, p. 48003, 2010.

\bibitem{kurz2011hegselmann}
S.~Kurz and J.~Rambau, ``On the {H}egselmann--{K}rause conjecture in opinion dynamics,'' \emph{Journal of Difference Equations and Applications}, vol.~17, no.~6, pp. 859--876, 2011.

\bibitem{sirbu2013cohesion}
A.~S{\^\i}rbu, V.~Loreto, V.~D. Servedio, and F.~Tria, ``Cohesion, consensus and extreme information in opinion dynamics,'' \emph{Advances in Complex Systems}, vol.~16, no.~06, p. 1350035, 2013.

\bibitem{sirbu2013opinion}
------, ``Opinion dynamics with disagreement and modulated information,'' \emph{Journal of Statistical Physics}, vol. 151, pp. 218--237, 2013.

\bibitem{li2020effect}
T.~Li and H.~Zhu, ``Effect of the media on the opinion dynamics in online social networks,'' \emph{Physica A: Statistical Mechanics and its Applications}, vol. 551, p. 124117, 2020.

\bibitem{yang2021opinion}
M.-H. Yang, J.-W. Yi, and L.~Chai, ``Opinion dynamics of the {DeGroot} model with rebels and advertising,'' in \emph{2021 China Automation Congress (CAC)}.\hskip 1em plus 0.5em minus 0.4em\relax IEEE, 2021, pp. 7493--7498.

\bibitem{muslim2024mass}
R.~Muslim, R.~A. Nqz, and M.~A. Khalif, ``Mass media and its impact on opinion dynamics of the nonlinear q-voter model,'' \emph{Physica A: Statistical Mechanics and its Applications}, vol. 633, p. 129358, 2024.

\bibitem{mandyam2012community}
S.~Mandyam and U.~Sridhar, ``Community learning from external information sources,'' in \emph{2012 IEEE/ACM International Conference on Advances in Social Networks Analysis and Mining}.\hskip 1em plus 0.5em minus 0.4em\relax IEEE, 2012, pp. 1329--1334.

\bibitem{mirtabatabaei2014eulerian}
A.~Mirtabatabaei, P.~Jia, and F.~Bullo, ``Eulerian opinion dynamics with bounded confidence and exogenous inputs,'' \emph{SIAM Journal on Applied Dynamical Systems}, vol.~13, no.~1, pp. 425--446, 2014.

\bibitem{quattrociocchi2014opinion}
W.~Quattrociocchi, G.~Caldarelli, and A.~Scala, ``Opinion dynamics on interacting networks: Media competition and social influence,'' \emph{Scientific reports}, vol.~4, no.~1, p. 4938, 2014.

\bibitem{mao2018spread}
Y.~Mao, S.~Bolouki, and E.~Akyol, ``Spread of information with confirmation bias in cyber-social networks,'' \emph{IEEE Transactions on Network Science and Engineering}, vol.~7, no.~2, pp. 688--700, 2018.

\bibitem{mao2022social}
Y.~Mao, N.~Hovakimyan, T.~Abdelzaher, and E.~Theodorou, ``Social system inference from noisy observations,'' \emph{IEEE Transactions on Computational Social Systems}, 2022.

\bibitem{gunducc2020effect}
S.~G{\"u}nd{\"u}{\c{c}}, ``The effect of social media on shaping individuals opinion formation,'' in \emph{Complex Networks and Their Applications VIII: Volume 2 Proceedings of the Eighth International Conference on Complex Networks and Their Applications COMPLEX NETWORKS 2019 8}.\hskip 1em plus 0.5em minus 0.4em\relax Springer, 2020, pp. 376--386.

\bibitem{galtung1965structure}
J.~Galtung and M.~H. Ruge, ``The structure of foreign news: The presentation of the {C}ongo, {C}uba and {C}yprus crises in four {N}orwegian newspapers,'' \emph{Journal of peace research}, vol.~2, no.~1, pp. 64--90, 1965.

\bibitem{glasauer2022individual}
S.~Glasauer and Z.~Shi, ``Individual beliefs about temporal continuity explain variation of perceptual biases,'' \emph{Scientific Reports}, vol.~12, no.~1, p. 10746, 2022.

\bibitem{andina2007reinforcement}
A.~Andina-D{\'\i}az, ``Reinforcement vs. change: The political influence of the media,'' \emph{Public Choice}, vol. 131, pp. 65--81, 2007.

\bibitem{bernhardt2008political}
D.~Bernhardt, S.~Krasa, and M.~Polborn, ``Political polarization and the electoral effects of media bias,'' \emph{Journal of Public Economics}, vol.~92, no. 5-6, pp. 1092--1104, 2008.

\bibitem{osborne1959brownian}
M.~F. Osborne, ``Brownian motion in the stock market,'' \emph{Operations research}, vol.~7, no.~2, pp. 145--173, 1959.

\bibitem{reddy2016simulating}
K.~Reddy and V.~Clinton, ``Simulating stock prices using geometric {B}rownian motion: Evidence from {A}ustralian companies,'' \emph{Australasian Accounting, Business and Finance Journal}, vol.~10, no.~3, pp. 23--47, 2016.

\bibitem{okawa2022predicting}
M.~Okawa and T.~Iwata, ``Predicting opinion dynamics via sociologically-informed neural networks,'' in \emph{Proceedings of the 28th ACM SIGKDD Conference on Knowledge Discovery and Data Mining}, 2022, pp. 1306--1316.

\bibitem{berger1981necessary}
R.~L. Berger, ``A necessary and sufficient condition for reaching a consensus using {DeGroot's} method,'' \emph{Journal of the American Statistical Association}, vol.~76, no. 374, pp. 415--418, 1981.

\bibitem{karatzas1991brownian}
I.~Karatzas and S.~Shreve, \emph{Brownian motion and stochastic calculus}.\hskip 1em plus 0.5em minus 0.4em\relax Springer Science \& Business Media, 1991, vol. 113.

\bibitem{bhattacharya2021random}
R.~N. Bhattacharya and E.~C. Waymire, \emph{Random walk, {B}rownian motion, and martingales}.\hskip 1em plus 0.5em minus 0.4em\relax Springer, 2021.

\bibitem{kailath1980linear}
T.~Kailath, \emph{Linear systems}.\hskip 1em plus 0.5em minus 0.4em\relax Prentice-Hall Englewood Cliffs, NJ, 1980, vol. 156.

\bibitem{horn2012matrix}
R.~A. Horn and C.~R. Johnson, \emph{Matrix analysis}.\hskip 1em plus 0.5em minus 0.4em\relax Cambridge university press, 2012.

\end{thebibliography}

\clearpage

\section*{Appendix}

\subsection{Theorem Proof}
\label{sec:proof}

\subsubsection{Theorem \ref{th:1}}
According to \cite{karatzas1991brownian}, we have Lemma \ref{le:1}.

\begin{lemma}\label{le:1}
    Consider the Brownian motion $\{x_t\}_{t\geqslant0}$ on $[0, 1]$, whose changing rate is equal to $c$. If the initial value $0<x_0<1$, then for $t>0$, $0<x<1$, the probability density functions of the first hitting time $T_0$ and $T_1$ are
    \begin{equation}
    \begin{cases}
        \displaystyle f_{T_0}(t,T_0<T_1|x_0)=\frac{1}{t}\sum_{n\in\mathbf{E}}(x_0-n)g(n,t|x_0),\\
        \displaystyle f_{T_1}(t,T_1<T_0|x_0)=\frac{1}{t}\sum_{n\in\mathbf{O}}(n-x_0)g(n,t|x_0).
    \end{cases}
    \label{eq:sT0}
    \end{equation}
    For $t>0$, $0<x<1$, the probability density function of $x_t$ is
    \begin{equation}
        f_{x_t}(x|x_0)=\sum_{n\in\mathbf{E}}g(n+x,t|x_0)-g(-n-x,t|x_0),
    \end{equation}
    where $g(x,t|x_0)=\frac{1}{\sqrt{2\pi c^2t}}\exp\left\{-\frac{(x-x_0)^2}{2c^2t}\right\}$.
\end{lemma}

From Lemma \ref{le:1}, \eqref{eq:pdf} is established. Note that the following random events are equivalent:
\begin{align}
    &\left\{\omega: s_{i,t}(\omega)=0\Big| s_{i,0}\right\}\notag\\
    =&\left\{\omega: T_0(\omega)\leqslant t,T_0(\omega)<T_1(\omega)\Big|s_{i,0}\right\}\notag\\
    =&\bigcup_{0\leqslant\tau\leqslant t}\left\{\omega: T_0(\omega)=\tau,T_0(\omega)<T_1(\omega)\Big|s_{i,0}\right\},\quad\text{and}\\
    &\left\{\omega: s_{i,t}(\omega)=1\Big| s_{i,0}\right\}\notag\\
    =&\left\{\omega: T_1(\omega)\leqslant t,T_1(\omega)<T_0(\omega)\Big|s_{i,0}\right\}\notag\\
    =&\bigcup_{0\leqslant\tau\leqslant t}\left\{\omega: T_1(\omega)=\tau,T_1(\omega)<T_0(\omega)\Big|s_{i,0}\right\}.
    \label{eq:T1u}
\end{align}
Thus, the probabilities of reaching the absorbing bounds are
\begin{align}
\mathbb{P}(s_{i,t}=0|s_{i,0}) &= \int_0^t f_{T_0}(\tau,T_0<T_1|s_{i,0})\dif \tau\notag\\
&= \int_0^t\frac{1}{\tau}\sum_{n\in\mathbf{E}}(s_{i,0}-n)g(n,\tau|s_{i,0})\dif \tau,\quad\text{and}\notag\\
\mathbb{P}(s_{i,t}=1|s_{i,0}) &= \int_0^t f_{T_1}(\tau,T_1<T_0|s_{i,0})\dif \tau\notag\\
&= \int_0^t\frac{1}{\tau}\sum_{n\in\mathbf{O}}(n-s_{i,0})g(n,\tau|s_{i,0})\dif \tau.
\end{align}
Here, we have established the proof of \eqref{equ:si=0}.

\subsubsection{Theorem \ref{co:1}}
As $t$ approaches infinity, the asymptotic transition probability function is
\begin{equation}
    \lim_{t\to\infty}g(x,t|x_0)\overset{\text{a.s.}}{=}0,\forall x\in(0,1).
\end{equation}
Therefore, from Theorem \ref{th:1}, the probability density function of $s_{i,t}$ is
\begin{equation}
    \lim_{t\to\infty}f_{s_{i,t}}(x,t|s_{i,0})\overset{\text{a.s.}}{=}0,\forall x\in(0,1).
\end{equation}
For $0\leqslant s\leqslant t$, we can prove that
\begin{align}
    \mathbb{E}(y_{i,t}|\mathscr{F}_s)&=s_{i,0}+c\mathbb{E}(z_{i,t}|\mathscr{F}_s)\notag\\
    &\overset{\text{a.s.}}{=}s_{i,0}+cz_{i,s}=y_{i,s},\quad\text{and}\\
    \mathbb{E}|y_{i,t}|&<\infty,
\end{align}
so $\{y_{i,t}\}_{t\geqslant0}$ is a continuous-time martingale. Because $T:=T_0\wedge T_1$ is a stopping time with respect to $\{y_{i,t}\}_{t\geqslant0}$, and from \cite{bhattacharya2021random} we have
\begin{align}
    &\mathbb{E}T<\infty,\quad\text{and}\\
    &\mathbb{E}\sup_{t\geqslant0}|y_{i,T\wedge t}|\leqslant1,
\end{align}
according to the Stopping Time Theorem, we can prove that
\begin{equation}
    \mathbb{E}(y_{i,T}|s_{i,0})=\mathbb{E}(y_{i,0}|s_{i,0})\overset{\text{a.s.}}{=}s_{i,0}.
\end{equation}
From the definition of mean, we have
\begin{equation}
    \mathbb{E}(y_{i,T}|s_{i,0})=\mathbb{P}(T_1<T_0|s_{i,0}),
\end{equation}
so
\begin{equation}
    \mathbb{P}(T_1<T_0|s_{i,0})\overset{\text{a.s.}}{=}s_{i,0}.
\end{equation}
From \eqref{eq:T1u}, note that the following random events are equivalent:
\begin{align}
    &\lim_{t\to\infty}\left\{\omega: s_{i,t}(\omega)=1\Big| s_{i,0}\right\}\notag\\
    =&\lim_{t\to\infty}\bigcup_{0\leqslant\tau\leqslant t}\left\{\omega: T_0(\omega)=\tau,T_1(\omega)<T_0(\omega)\Big|s_{i,0}\right\}\notag\\
    =&\left\{\omega:T_1(\omega)<T_0(\omega)\Big|s_{i,0}\right\}.
\end{align}
Because
\begin{equation}
    \left\{\omega: s_{i,\sigma}(\omega)=1\Big| s_{i,0}\right\}\subset\left\{\omega: s_{i,\tau}(\omega)=1\Big| s_{i,0}\right\}, \forall\sigma<\tau,
\end{equation}
according to the continuity of probability, we have
\begin{align}
    \lim_{t\to\infty}\mathbb{P}(s_{i,t}=1|s_{i,0})&=\mathbb{P}\left(\lim_{t\to\infty}s_{i,t}\notag=1\Big|s_{i,0}\right)\\
    &=\mathbb{P}(T_1<T_0|s_{i,0})\overset{\text{a.s.}}{=}s_{i,0}.
\end{align}
Thus, we have
\begin{equation}
    \lim_{t\to\infty}\mathbb{P}(s_{i,t}=0|s_{i,0})\overset{\text{a.s.}}{=}1-s_{i,0}.
\end{equation}
Here, we have established the proof of \eqref{eq:asy}.

\subsubsection{Theorem \ref{th:2}}
According to \cite{bhattacharya2021random}, we have Lemma \ref{le:2}.
\begin{lemma}\label{le:2}
    If $\{x_t\}_{t\geqslant0}$ is a martingale and $T$ is a stopping time, then the stop process $\{x_{T\wedge t}\}_{t\geqslant0}$ is also a martingale.
\end{lemma}

From Lemma \ref{le:2}, since $\{y_{i,t}\}_{t\geqslant0}$ is a martingale and $T$ is a stopping time, the stopped process $\{s_{i,t}\}_{t\geqslant0}=\{y_{i,T\wedge t}\}_{t\geqslant0}$ is a martingale. Therefore, we have
\begin{equation}
    \mathbb{E}(s_{i,t}|s_{i,0})\overset{\text{a.s.}}{=}s_{i,0}.
\end{equation}
According to the Law of Total Mean, we have
\begin{equation}
    \mathbb{E}s_{i,t}=\mathbb{E}[\mathbb{E}(s_{i,t}|s_{i,0})]=\mathbb{E}s_{i,0}=\mu.
\end{equation}

\subsubsection{Theorem \ref{th:3}}
When $t$ is small, we can approximate the Brownian motion with absorbing bounds as the traditional Brownian motion, so
\begin{equation}
    \mathbb{D}(s_{i,t}|s_{i,0})=\mathbb{D}(y_{i,T\wedge t}|s_{i,0})=c^2\mathbb{D}z_{i,T\wedge t}\leqslant c^2t.
    \label{eq:b1}
\end{equation}
From Theorem \ref{co:1}, as $t$ approaches infinity, the message almost surely hits the absorbing bounds, so the variance is
\begin{align}
    \mathbb{D}(s_{i,t}|s_{i,0})&\overset{\text{a.s.}}{\leqslant}\mathbb{P}(s_{i,t}=0|s_{i,0})s_{i,0}^2+\mathbb{P}(s_{i,t}=1|s_{i,0})(1-s_{i,0})^2\notag\\
    &=(1-s_{i,0})s_{i,0}^2+s_{i,0}(1-s_{i,0})^2\notag\\
    &=s_{i,0}(1-s_{i,0}).
    \label{eq:b2}
\end{align}
From \eqref{eq:b1} and \eqref{eq:b2}, we can establish the upper bound of the variance:
\begin{equation}
    \mathbb{D}(s_{i,t}|s_{i,0})\overset{\text{a.s.}}{\leqslant}(c^2t)\wedge[s_{i,0}(1-s_{i,0})].
\end{equation}
According to the Law of Total Variance, when $t$ is small, the variance is
\begin{align}
    \mathbb{D}s_{i,t}&=\mathbb{E}[\mathbb{D}(s_{i,t}|s_{i,0})]+\mathbb{D}[\mathbb{E}(s_{i,t}|s_{i,0})]\notag\\
    &=c^2t+\mathbb{D}s_{i,0}=c^2t+\delta^2.
\end{align}
As $t$ approaches infinity, the variance is
\begin{align}
    \lim_{t\to\infty}\mathbb{D}s_{i,t}&=\mathbb{E}[s_{i,0}(1-s_{i,0})]+\delta^2=\mathbb{E}s_{i,0}-\mathbb{E}(s_{i,0})^2+\delta^2\notag\\
    &=\mu-(\mu^2+\delta^2)+\delta^2=\mu(1-\mu).
\end{align}

\subsubsection{Theorem \ref{th:4}}
For the convenience of expression, we denote $\boldsymbol{A}:=\alpha\boldsymbol{W}-\boldsymbol{I}$ and $\boldsymbol{B}:=(1-\alpha)\boldsymbol{U}$ hereinafter. According to \cite{kailath1980linear}, we have Lemma \ref{le:3}.

\begin{lemma}\label{le:3}
The solution of the matrix differential equation $\dot{\boldsymbol{x}}_t=\boldsymbol{A}\boldsymbol{x}_t+\boldsymbol{B}\boldsymbol{u}_t$ is
\begin{equation}
    \boldsymbol{x}_t=\mathrm{e}^{\boldsymbol{A}t}\boldsymbol{x}_0+\int_0^t\mathrm{e}^{\boldsymbol{A}(t-\tau)}\boldsymbol{B}\boldsymbol{u}_\tau\dif\tau.
\end{equation}
\end{lemma}

The MED model is \eqref{eq:med}. According to Lemma \ref{le:3}, Theorem \ref{th:4} is established obviously.

\subsubsection{Theorem \ref{co:4}}
From Theorem \ref{th:4}, the mean of agents' opinions is
\begin{align}
    \mathbb{E}\boldsymbol{o}_t&=\mathbb{E}\left[\mathrm{e}^{\boldsymbol{A}t}\boldsymbol{o}_0+\int_0^t\mathrm{e}^{\boldsymbol{A}(t-\tau)}\boldsymbol{B}\boldsymbol{s}_\tau\dif\tau\right]\notag\\
    &=\mathrm{e}^{\boldsymbol{A}t}\boldsymbol{o}_0+\int_0^t\mathrm{e}^{\boldsymbol{A}(t-\tau)}\boldsymbol{B}\mathbb{E}\boldsymbol{s}_\tau\dif\tau\notag\\
    &=\mathrm{e}^{\boldsymbol{A}t}\boldsymbol{o}_0+\mu\int_0^t\mathrm{e}^{\boldsymbol{A}(t-\tau)}\dif\tau\boldsymbol{B}\boldsymbol{1}\notag\\
    &=\mathrm{e}^{\boldsymbol{A}t}\boldsymbol{o}_0+\mu(1-\alpha)(\mathrm{e}^{\boldsymbol{A}t}-\boldsymbol{I})\boldsymbol{A}^{-1}\boldsymbol{1}.
    \label{eq:eot1}
\end{align}

From \cite{horn2012matrix}, because $\boldsymbol{1}$ is an eigenvector of the row-stochastic matrix $\boldsymbol{W}$ with eigenvalue $1$, it is also an eigenvector of $\boldsymbol{A}$ with eigenvalue $\alpha-1$. Furthermore, from \cite{horn2012matrix}, because $\boldsymbol{W}$ is a row-stochastic matrix, the moduli of all eigenvalues are less than or equal to $1$. So the moduli of all eigenvalues of matrix $\boldsymbol{A}$ are less than or equal to $\alpha-1<0$. That is, $\boldsymbol{A}$ is nonsingular, and $\boldsymbol{1}$ is an eigenvector of $\boldsymbol{A}^{-1}$ with eigenvalue $(\alpha-1)^{-1}$, i.e.,
\begin{equation}
    \boldsymbol{A}^{-1}\boldsymbol{1}=(\alpha-1)^{-1}\boldsymbol{1}.
    \label{eq:A1}
\end{equation}
Therefore, from \eqref{eq:eot1} and \eqref{eq:A1}, we have \eqref{eq:Eo(t)}. 

Furthermore, because the moduli of all eigenvalues of matrix $\boldsymbol{A}$ are negative, as $t$ approaches infinity, the matrix exponential function $\mathrm{e}^{\boldsymbol{A}t}$ tends to the zero matrix. Therefore, we have
\begin{align}
    \lim_{t\to\infty}\mathbb{E}\boldsymbol{o}_t&=\lim_{t\to\infty}\mathrm{e}^{\boldsymbol{A}t}\boldsymbol{o}_0+\mu(1-\alpha)\boldsymbol{A}^{-1}(\mathrm{e}^{\boldsymbol{A}t}-\boldsymbol{I})\boldsymbol{1}\notag\\
    &=\mu(\alpha-1)\boldsymbol{A}^{-1}\boldsymbol{1}=\mu\boldsymbol{1}.
\end{align}

\subsubsection{Theorem \ref{co:5}}
From Theorem \ref{th:4}, the variance of agents' opinions is
    \begin{align}
        \mathbb{D}\boldsymbol{o}_t&=\mathbb{D}\left[\mathrm{e}^{\boldsymbol{A}t}\boldsymbol{o}_0+\int_0^t\mathrm{e}^{\boldsymbol{A}(t-\tau)}\boldsymbol{B}\boldsymbol{s}_\tau\dif\tau\right]\notag\\
        &=\mathbb{D}\left[\int_0^t\mathrm{e}^{\boldsymbol{A}(t-\tau)}\boldsymbol{B}\boldsymbol{s}_\tau\dif\tau\right]\notag\\
        &=\mathrm{diag}\int_0^t\int_0^t\mathrm{e}^{\boldsymbol{A}(t-\sigma)}\boldsymbol{B}\boldsymbol{\Sigma}_{\sigma,\tau}\boldsymbol{B}^\top \mathrm{e}^{\boldsymbol{A}^\top(t-\tau)}\dif \sigma\dif \tau.
    \end{align}

As $t$ approaches infinity, we have $\lim_{t\to\infty}\dot{\boldsymbol{o}}_t\overset{\text{a.s.}}{=}\boldsymbol{0}$. By setting both sides of the MED model \eqref{eq:med} to zero vectors, we have
    \begin{equation}
        \lim_{t\to\infty}\boldsymbol{o}_t\overset{\text{a.s.}}{=}\lim_{t\to\infty}\boldsymbol{A}^{-1}\boldsymbol{B}\boldsymbol{s}_t.
    \end{equation}
    
    As $t$ approaches infinity, from Theorem \ref{th:1}, each element $s_{i,t}$ in $\boldsymbol{s}_t$ adheres to a binomial distribution, with a probability of $s_{i,0}$ for taking the value of $1$, and a probability of $1-s_{i,0}$ for taking the value of $0$. Therefore, the variance is
    \begin{align}
        \mathbb{D}\boldsymbol{o}_t&=\mathrm{diag}(\boldsymbol{A}^{-1}\boldsymbol{B}\boldsymbol{\Sigma}_{t,t}\boldsymbol{B}^\top\boldsymbol{A}^{-\top})\notag\\
        &=\mu(1-\mu)\mathrm{diag}(\boldsymbol{A}^{-1}\boldsymbol{B}\boldsymbol{B}^\top\boldsymbol{A}^{-\top}).
    \end{align}
    
    Here, we use Theorem \ref{th:3} to calculate the covariance. The diagonal elements of $\boldsymbol{\Sigma}_{t,t}$ are equal to the variance as shown in \eqref{eq:vars}, while the off-diagonal elements are zero because $\{z_{i,t}\}_{t\geqslant0}$ are identical for $i\in\mathscr{M}$, i.e.,
    \begin{equation}
        \boldsymbol{\Sigma}_{t,t}=\mu(1-\mu)\boldsymbol{I}.
    \end{equation}

\end{document}